\documentstyle[opex]{article}

\newcommand{\scrout}{\mbox{\scriptsize\rm out}}

\begin{document}
\title{Sampling quantum phase space with~squeezed states}

\author{Konrad Banaszek$^{\ast}$ and Krzysztof W\'odkiewicz}
\address{Instytut Fizyki Teoretycznej, Uniwersytet Warszawski,
Ho\.{z}a 69, PL-00-681~Warszawa, Poland}
\email{$^\ast$E-mail address: Konrad.Banaszek@fuw.edu.pl}
\begin{abstract}
We study the application of squeezed states in a quantum optical scheme
for direct sampling of the phase space by photon counting.  We prove
that the detection setup with a squeezed coherent probe field is
equivalent to the probing of the squeezed signal field with a coherent
state.  An example of the Schr\"{o}dinger cat state measurement shows
that the use of squeezed states allows one to detect clearly the
interference between distinct phase space components despite losses
through the unused output port of the setup.
\end{abstract}

\begin{OEReferences}
\item\label{CahiGlauPR69}
K.~E.~Cahill and R.~J.~Glauber, ``Density operators and 
quasiprobability distributions,'' Phys.\ Rev.\  
{\bf 177}, 1882--1902 (1969).

\item\label{SmitBeckPRL93}
D.~T.~Smithey, M.~Beck, M.~G.~Raymer, and A.~Faridani,
``Measurement of the Wigner distribution and the density matrix
of a light mode using optical homodyne tomography: Application
to squeezed states and the vacuum,'' \prl {\bf 70}, 1244--1247
(1993).

\item\label{WallVogePRA96}
S. Wallentowitz and W. Vogel, ``Unbalanced homodyning for quantum state
measurements,'' \pra {\bf 53}, 4528--4533 (1996).

\item\label{BanaWodkPRL96}
K. Banaszek and K. W\'{o}dkiewicz, ``Direct sampling of quantum
phase space by photon counting,'' \prl {\bf 76}, 4344--4347 (1996).

\item\label{LoudKnigJMO87}
R.~Loudon and P.~L.~Knight, ``Squeezed light,'' \jmo {\bf 34},
709--759 (1987).
 
\item\label{LeonPaulPRL94}
U. Leonhardt and H. Paul,
``High-accuracy optical homodyne detection with low-efficiency
detectors: `Preamplification' from antisqueezing'', \prl
{\bf 72}, 4086--4089 (1994).

\item\label{LeonPaulPRA93}
U. Leonhardt and H. Paul, 
``Realistic  optical homodyne measurements and quasidistribution functions,''
\pra {\bf 48}, 4598--4604 (1993).

\item\label{BanaWodkPRA97}
K. Banaszek and K. W\'{o}dkiewicz,
``Operational theory of homodyne detection,''
\pra {\bf 55}, 3117--3123 (1997).

\item\label{SchlPernPRA91}
W. Schleich, M. Pernigo, and F. LeKien, ``Nonclassical state
from two pseudoclassical states,'' \pra {\bf 44}, 2172--2187 (1991).

\item\label{BuzeKnigPiO95}
V. Bu\v{z}ek and P. L. Knight, ``Quantum interference, superposition
states of light, and nonclassical effects,'' in {\it Progress in
Optics XXXIV}, ed.\ by E. Wolf (north-Holland, Amsterdam, 1995),
1--158.

\end{OEReferences}

\section{Introduction}
Phase space quasidistribution functions are a convenient way of
characterizing the quantum state of optical radiation
[\ref{CahiGlauPR69}]. Over past
several years, they have gained experimental significance due to the
reconstruction of the Wigner function of a single light mode performed using
tomographic algorithms
[\ref{SmitBeckPRL93}]. Recently, an alternative method for measuring
quasidistribution functions of a light mode has been proposed
[\ref{WallVogePRA96},\ref{BanaWodkPRL96}]. The method is
based on photon counting of the signal field superposed on a probe
field in a 
coherent state. The advantage of this method is that there is no
complicated numerical processing of the experimental data. A simple
arithmetic operation performed on the photocount statistics yields
directly the value of a quasidistribution function at a point defined
by the amplitude and the phase of the coherent field.


The purpose of this communication
is to study the application of squeezed
states in the proposed photon counting scheme. The most important feature
of squeezed states is that quantum fluctuations in some observables
are reduced below the coherent state level [\ref{LoudKnigJMO87}]. 
In the context of optical
homodyne tomography, the squeezing transformation has been shown to be
capable of compensating for the deleterious effect of low detection
efficiency [\ref{LeonPaulPRL94}].
Therefore, it is interesting to discuss the information on
the quantum state of light which can be retrieved in a photon counting
experiment using squeezed states.

\section{Experimental scheme}
We start with a brief description of the proposed setup, depicted in
Fig.~1. The field incident on a photodetector is a
combination, performed using a beam splitter with a power transmission
$T$, of a transmitted signal mode and a reflected probe mode. The
statistics of the detector counts $\{p_n\}$ is used to calculate an
alternating series $\sum_{n=0}^{\infty} (-1)^n p_n$. In terms
of the outgoing mode, this series is given by the expectation value of
the parity operator:
\begin{equation}
\hat{\Pi} = 
(-1)^{\hat{a}_{\mbox{\tiny\rm out}}^{\dagger} 
\hat{a}_{\mbox{\tiny\rm out}}}, 
\end{equation}
where the annihilation operator of the
outgoing mode 
$\hat{a}_{\scrout}$
is a linear combination of the
signal and the probe field operators:
\begin{equation}
\hat{a}_{\scrout} = \sqrt{T} \hat{a}_{S} - \sqrt{1-T}
\hat{a}_{P}.
\end{equation}

The expectation value of the measured observable involves statistical
properties of both the signal and the probe modes. The operator
$\hat{\Pi}$ can be written in the following normally ordered form:
\begin{equation}
\label{Eq:NormOrd}
\hat{\Pi} = \; : \exp[-2
(\sqrt{T}\hat{a}_{S}^{\dagger} - \sqrt{1-T} \hat{a}_{P}^{\dagger})
(\sqrt{T}\hat{a}_{S} - \sqrt{1-T} \hat{a}_{P})]:,
\end{equation}
which has a clear and intuitive interpretation within the Wigner
function formalism: the measured quantity is proportional to the phase
space integral of the product of the signal and the probe Wigner
functions with relatively rescaled parameterizations [\ref{BanaWodkPRL96}].
 Hence the proposed scheme is a realization of direct sampling of
the quantum phase space.

An important class of probe fields are coherent states $\hat{a}_S
|\alpha\rangle_P = \alpha | \alpha \rangle_P$. 
The quantum expectation value over the probe mode can be easily
evaluated in this case
using the normally 
ordered form given in Eq.~(\ref{Eq:NormOrd}). Thus the measured
observable is given by the following 
operator acting in the Hilbert space
of the signal mode:
\begin{equation}
\langle \alpha | \hat{\Pi} | \alpha \rangle_P = \;
:\exp[ -2 (\sqrt{T} \hat{a}_{S}^{\dagger} - \sqrt{1-T}
\alpha^{\ast})(\sqrt{T} \hat{a}_{S} - \sqrt{1-T} \alpha)]:.
\end{equation}
This observable is closely related to a certain quasidistribution
function. The most straightforward way to identify this link
is to recall that an $s$-ordered quasidistribution
function at a complex phase space point $\beta$ is given by the 
expectation value of the normally ordered operator:
\begin{equation}
\hat{U}(\beta;s) = \frac{2}{\pi(1-s)} : \exp \left[
- \frac{2}{1-s} (\hat{a}_S^{\dagger} - \beta^{\ast})
(\hat{a}_S - \beta) \right]:.
\end{equation}
After a simple rearrangement of parameters we finally arrive at the
formula:
\begin{equation}
\langle \alpha | \hat{\Pi} | \alpha \rangle_P =
\frac{\pi}{2T} \hat{U} \left( \sqrt{\frac{1-T}{T}} \alpha ;
- \frac{1-T}{T} \right).
\end{equation}
Thus, the alternating series computed from the photocount statistics
yields the value of a quasidistribution function at a point
$\sqrt{(1-T)/T}\alpha$ defined by the amplitude and the phase of the
probe coherent field. The complete quasidistribution function can be
scanned point--by--point by changing the probe field parameters.

The ordering of the measured quasidistribution function depends on the
beam splitter transmission. This is a consequence of the fact that a
fraction of the signal field escapes through the second unused output
port of the beam splitter. These losses of the field lower the ordering
of the detected observable. This effect is analogous to the one
appearing in balanced homodyne detection with imperfect detectors
[\ref{LeonPaulPRA93},\ref{BanaWodkPRA97}]. In the limit $T\rightarrow
1$, when the complete signal field is detected, we measure directly the
Wigner function, corresponding to the symmetric ordering.

\begin{figure}
\centerline{\epsfbox{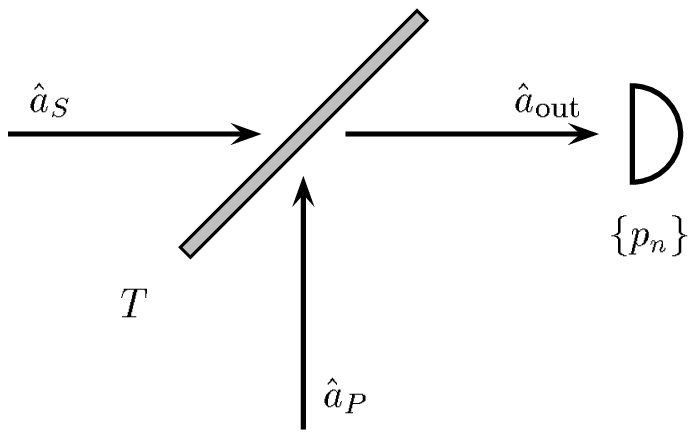}}
\vspace{10pt}
\begin{center}
\begin{minipage}{4.25in}
\footnotesize Fig.\ 1. The setup for direct probing of the
quantum phase space. 
The detector measures
the photocount statistics $\{p_n\}$ of a signal
$\hat{a}_{S}$ combined
with a probe field $\hat{a}_{P}$ using a beam splitter with
a power transmission~$T$.
\end{minipage}
\end{center}
\end{figure}
 
\section{Sampling with squeezed state}
We will now consider the case when a squeezed coherent state
$S_{P}(r,\varphi)
|\alpha\rangle_{P}$ enters through the probe port of the beam splitter.
We use the following definition of the squeezing operator for
an $i$th mode: 
\begin{equation} 
S_{i}(r,\varphi)
= \exp[r(
e^{-i\varphi}\hat{a}_{i}^2
-
e^{i\varphi}(\hat{a}_{i}^\dagger)^2)/2].
\end{equation} 
The detected quantity is now 
given by the expectation value of the following
operator acting in the Hilbert space of the signal mode:
\begin{equation}
\langle\hat{\Pi}\rangle_P = \langle \alpha | \hat{S}^{\dagger}_{P}
(r,\varphi) \hat{\Pi} \hat{S}_{P}(r,\varphi) | \alpha \rangle_{P}.
\end{equation}

In order to find an interpretation for 
this observable, we will derive a formula for
the squeezing transformations of the parity operator $\hat{\Pi}$.
We start from a simple unitary transformation:
\begin{equation}
(-1)^{\hat{a}^{\dagger}_{\mbox{\tiny\rm out}}
\hat{a}_{\mbox{\tiny\rm out}}}
\hat{a}_{\scrout}^{2}
(-1)^{\hat{a}^{\dagger}_{\mbox{\tiny\rm out}}
\hat{a}_{\mbox{\tiny\rm out}}}
=
e^{i\pi \hat{a}^{\dagger}_{\mbox{\tiny\rm out}}
\hat{a}_{\mbox{\tiny\rm out}}}
\hat{a}_{\scrout}^{2}
e^{-i\pi \hat{a}^{\dagger}_{\mbox{\tiny\rm out}}
\hat{a}_{\mbox{\tiny\rm out}}}
=
e^{-2\pi i}\hat{a}_{\scrout}^{2}
=
\hat{a}_{\scrout}^{2}.
\end{equation}
This equation implies the commutator:
\begin{equation}
[(-1)^{\hat{a}^{\dagger}_{\mbox{\tiny\rm out}}
\hat{a}_{\mbox{\tiny\rm out}}},
e^{i\varphi} (\hat{a}_{\scrout}^{\dagger})^2 - 
e^{-i\varphi} \hat{a}_{\scrout}^2] =0,
\end{equation}
which states that generation or annihilation of pairs of
photons
conserves parity. Therefore, the parity operator is invariant
under the squeezing transformation:
\begin{equation}
\hat{S}_{\scrout}^{\dagger}(r,\varphi) \hat{\Pi}
\hat{S}_{\scrout}(r,\varphi) = \hat{\Pi}.
\end{equation}
This identity has nontrivial consequences
 when written in terms of the signal and
the probe modes. It is equivalent to the equation:
\begin{equation}
\hat{S}^{\dagger}_{S}(r,\varphi)
\hat{S}^{\dagger}_{P}(r,\varphi)
\hat{\Pi}
\hat{S}_{P}(r,\varphi)
\hat{S}_{S}(r,\varphi)
= \hat{\Pi}
\end{equation}
which, after moving the signal squeezing operators to the right hand
side,
yields the following result:
\begin{equation}
\label{Eq:Przeniesienie}
\hat{S}^{\dagger}_{P}(r,\varphi)
\hat{\Pi}
\hat{S}_{P}(r,\varphi)
=
\hat{S}^{\dagger}_{S}(-r,\varphi)
\hat{\Pi}
\hat{S}_{S}(-r,\varphi)
\end{equation}
This formula shows that squeezing of the probe mode is equivalent 
to squeezing of the signal mode with the opposite sign of the parameter
$r$. This change of the sign swaps the field quadratures that get squeezed
or antisqueezed under the squeezing transformation.

Finally we obtain the following explicit expression for the detected
signal field observable:
\begin{eqnarray}
\langle\hat{\Pi}\rangle_{P} & = & \hat{S}^{\dagger}_{S}(-r,\varphi)
\langle\alpha|\hat{\Pi}|\alpha\rangle_P \hat{S}_{S}(-r,\varphi)
\nonumber \\
\label{Eq:ExpPiSq}
& = & \frac{\pi}{2T}  \;
\hat{S}^{\dagger}_{S}(-r,\varphi) \;
\hat{U}
\left(
\sqrt{\frac{1-T}{T}}\alpha ; - \frac{1-T}{T}
\right)
\hat{S}_{S}(-r,\varphi).
\end{eqnarray}
Thus, the setup delivers again an $s=-(1-T)/T$-ordered
quasidistribution function at a phase space point $\sqrt{(1-T)/T}$, but
corresponding to a {\em squeezed} signal field.

Let us note that it was possible to carry the squeezing transformation
from the probe to the signal degree of freedom only due to a specific
form of the measured observable. We have explicitly used the
conservation of the parity operator during generation or annihilation
of pairs of photons. For a general observable defined for the outgoing
mode $\hat{a}_{\scrout}$, 
there is no formula analogous to Eq.~(\ref{Eq:Przeniesienie}).

\section{Detection of Schr\"{o}dinger cat state}

As an illustration, we will consider a photon counting experiment for a
Schr\"{o}dinger cat state, which is a 
quantum superposition of two coherent states [\ref{SchlPernPRA91}]:
\begin{equation}
|\psi\rangle = \frac{|i\kappa\rangle + | -i\kappa\rangle}%
{\sqrt{2+ 2\exp(-2\kappa^2)}},
\end{equation}
where $\kappa$ is a real parameter.
The Wigner function of such a state contains, in addition to two positive
peaks corresponding to the coherent states, an oscillating term
originating from quantum interference between the classical--like
components. This nonclassical feature is extremely fragile, and
disappears very quickly in the presence of dissipation
[\ref{BuzeKnigPiO95}]. 

As we have found in Eq.~(\ref{Eq:ExpPiSq}), 
the outcome of the photon counting experiment
with a squeezed probe field is related to 
an $s$-ordered quasidistribution of the squeezed
Schr\"{o}dinger cat state $\hat{S}_{S}(-r,\varphi)|\psi\rangle$.
For simplicity, we will restrict ourselves to the case $\varphi=0$.
A simple but lengthy calculation yields the explicit formula for
the phase space quasidistribution at a complex point $\beta = q+ip$:
\begin{eqnarray}
\lefteqn{
\langle\psi | \hat{S}_{S}^{\dagger}(-r,0)
\hat{U}(q+ip; s) \hat{S}_{S}(-r,0) | \psi\rangle
} & & \nonumber \\
& = &
\frac{\displaystyle\exp\left(-\frac{2q^2}{e^{2r}-s}\right)}%
{\pi[1+\exp(-2\kappa^2)]\sqrt{1-2s\cosh 2r +s^2}}
\left\{
 \exp\left[-\frac{2(p-e^{-r}\kappa)^2}{e^{-2r}-s}\right]
\right.
\nonumber \\
& &
\left.
+\exp\left[-\frac{2(p+e^{-r}\kappa)^2}{e^{-2r}-s}\right]
+ 2 \exp\left( \frac{2s\kappa^2}{e^{2r}-s}
- \frac{2p^2}{e^{-2r}-s}\right) 
\cos\left(\frac{4e^{r}\kappa q}{e^{2r}-s}\right)
\right\}.
\nonumber \\
& &
\end{eqnarray}

In Fig.~2 we depict the expectation value of the parity operator
$\langle\hat{\Pi}\rangle$ as a function of the rescaled complex probe
field amplitude $\beta = \sqrt{(1-T)/T}\alpha$. For comparison, we show
two cases: when the Schr\"{o}digner cat state is probed with coherent
states $|\alpha\rangle_P$ and squeezed coherent states
$\hat{S}_{P}(r=1,0)|\alpha\rangle_{P}$. The beam splitter transmission
is $T=80\%$. When coherent states are used, only a faint trace of the
oscillatory pattern can be noticed due to losses of the signal field.
In contrast, probing of the
Schr\"{o}dinger cat state with suitably chosen squeezed states
yields a
clear picture of quantum coherence between distinct phase space
components. This effect is particularly surprising if we realize
that $20\%$ of the signal field power is lost through the unused output
port of the beam splitter.

The visibility of the oscillatory pattern depends substantially on the
sign of the squeezing parameter $r$. This can be most easily
understood using the Wigner phase space description of the discussed
scheme [\ref{BanaWodkPRL96}]. In order to detect the interference,
fluctuations in the probe squeezed states have to be reduced in the
direction corresponding to the rapid oscillations of the Wigner
function
corresponding to the Schr\"{o}dinger cat state.  The width of the
rescaled probe Wigner function along the squeezed direction must be smaller
than the spacing between the interference fringes.

\section{Conclusions}
We have studied the quantum optical scheme for direct sampling of
the quantum phase space using squeezed coherent states. We have shown
that 
squeezing transformations performed on the signal and the probe input
ports of the setup are equivalent. The application of  squeezed states with
the appropriately chosen squeezing direction allows one to detect quantum
interference despite losses through the unused output port of the
setup.

\section*{Acknowledgements}
This work has been
partially supported by the Polish KBN grant 2 PO3B~006~11. K.B. would like
to acknowledge fruitful discussions with E.~Czuchry.

\begin{figure}
\begin{center}
\setlength{\unitlength}{1.in}
\begin{picture}(4,6)
\put(0,0){\epsfxsize=4in\epsfbox{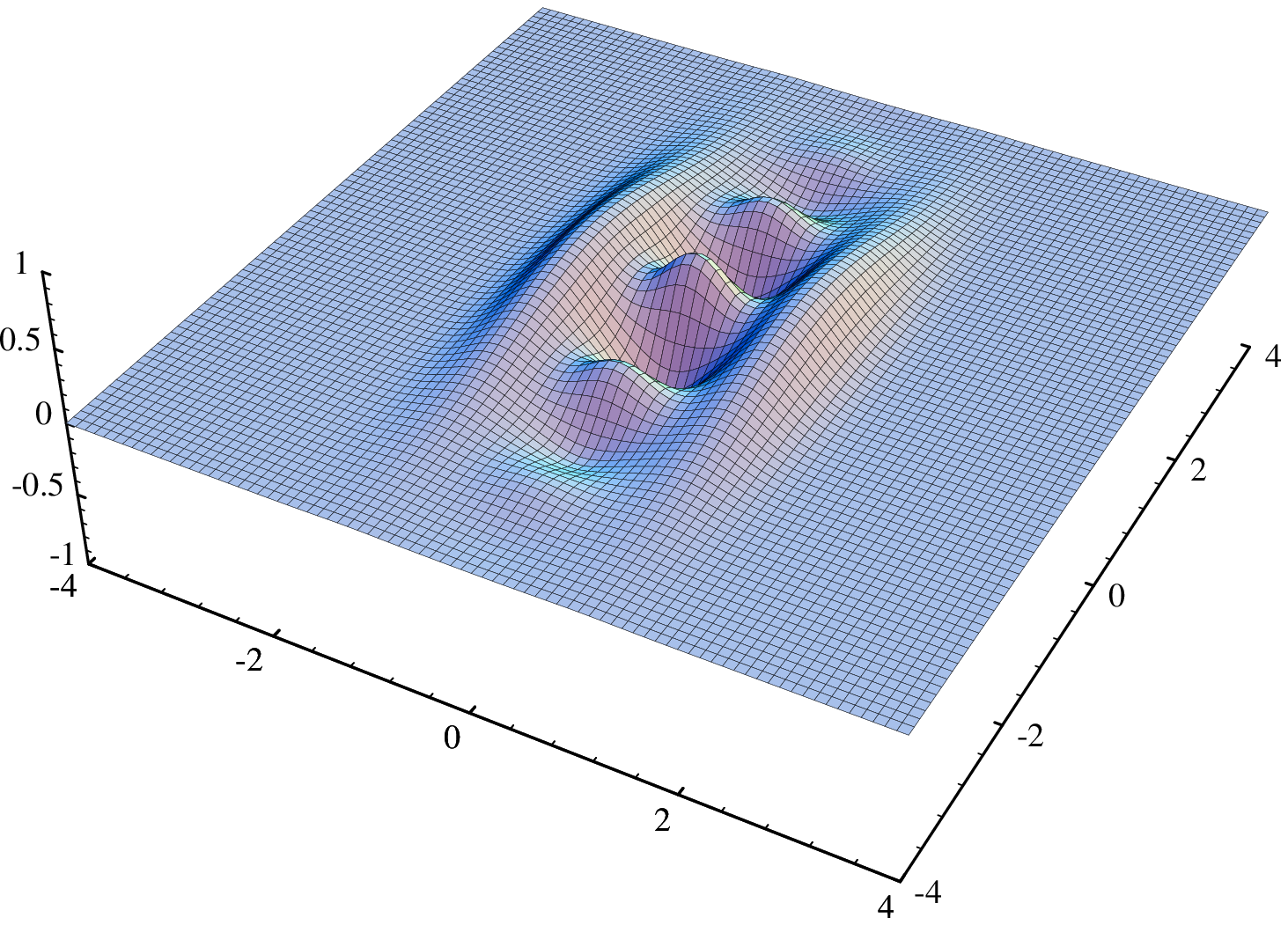}}
\put(4,2.8){\makebox(0,0)[rt]{\large (b)}}
\put(1.3,0.3){\makebox(0,0){Im$\beta$}}
\put(3.8,0.8){\makebox(0,0){Re$\beta$}}
\put(0.1,2.3){\makebox(0,0){$\langle\hat{\Pi}\rangle$}}
\put(0,3){\epsfxsize=4in\epsfbox{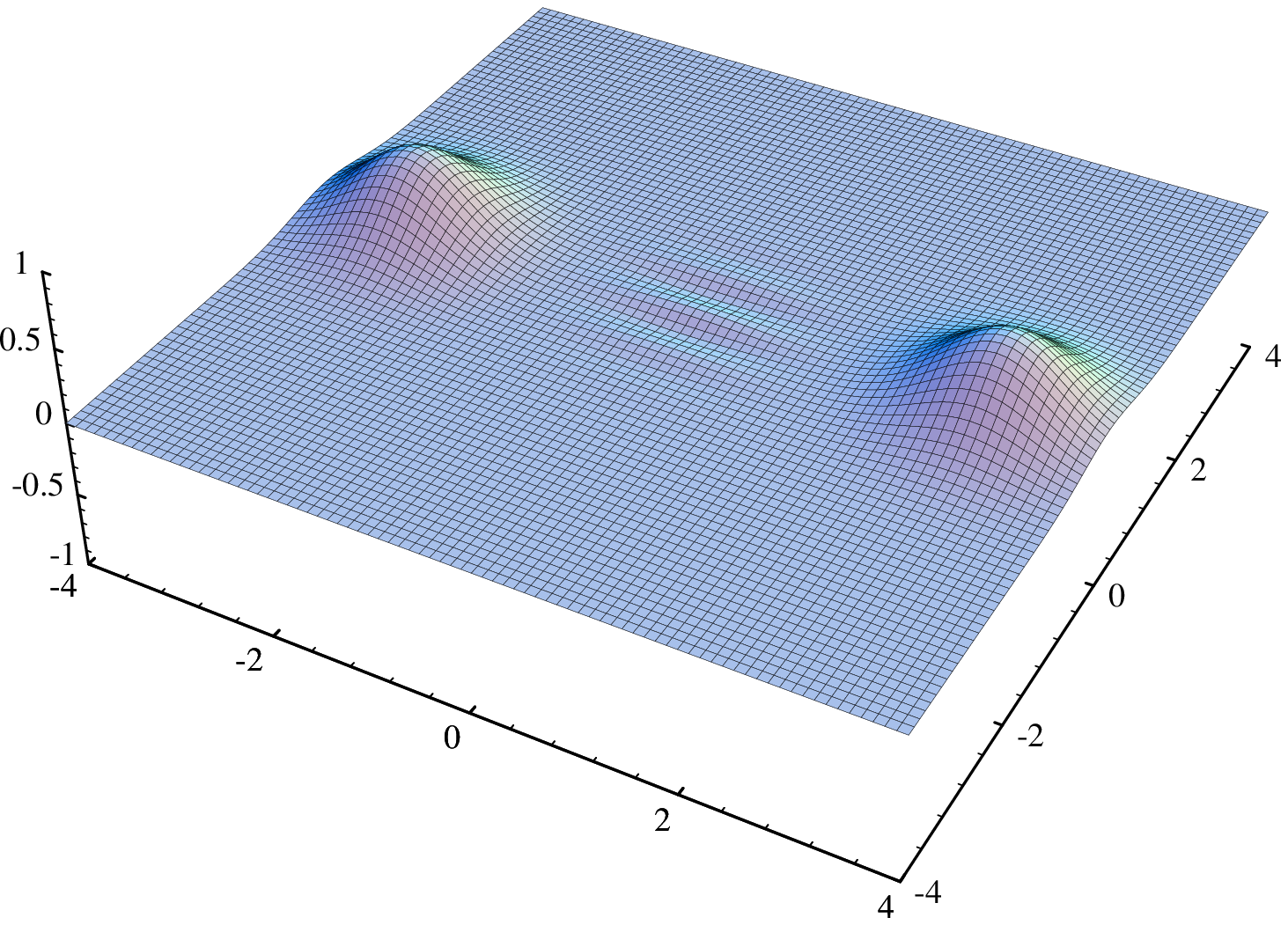}}
\put(4,5.8){\makebox(0,0)[rt]{\large (a)}}
\put(1.3,3.3){\makebox(0,0){Im$\beta$}}
\put(3.8,3.8){\makebox(0,0){Re$\beta$}}
\put(0.1,5.3){\makebox(0,0){$\langle\hat{\Pi}\rangle$}}
\end{picture}
\end{center}
\vspace{10pt}
\begin{center}
\begin{minipage}{4.25in}
\footnotesize
Fig.\ 2.
Sampling the Schr\"{o}digner cat state
$|\psi\rangle \propto |3i\rangle + |-3i\rangle$ with: (a) coherent
states $|\alpha\rangle_P$ and (b) squeezed states $\hat{S}_{P}
(r=1,0)|\alpha\rangle_{P}$. The plots show
the expectation value of the parity operator $\langle\hat{\Pi}\rangle$
as a function of the rescaled complex probe field amplitude $\beta
= \sqrt{(1-T)/T}\alpha$.
The beam splitter transmission is $T=80\%$.
\end{minipage}
\end{center}
\end{figure}
 
\end{document}